\documentclass[sigconf, nonacm]{acmart}

\usepackage{amsmath}
\usepackage{algorithm}
\usepackage{algpseudocode}
\usepackage{multirow} % 表のセル結合用
\usepackage{tabularx} % 表の幅を調整するため
\usepackage{tcolorbox}

\makeatletter
\def\blfootnote{\xdef\@thefnmark{}\@footnotetext}
\makeatother

\begin{document}

% \title{A Tuning-Free Recommendation Framework for Text-Rich Industrial Domains}
\title{LLMAR: A Tuning-Free Recommendation Framework for Sparse and Text-Rich Industrial Domains}

\author{Ryogo Hishikawa}
\email{ryogo.hishikawa.xq@hitachi.com}
\affiliation{%
  \institution{Hitachi, Ltd. Research and Development Group}
  \city{Hitachi}
  \state{Ibaraki}
  \country{Japan}
}

\author{Ichiro Kataoka}
\email{ichiro.kataoka.vf@hitachi.com}
\affiliation{%
  \institution{Hitachi, Ltd. Research and Development Group}
  \city{Hitachi}
  \state{Ibaraki}
  \country{Japan}
}

\author{Shinya Yuda}
\email{shinya.yuda.cv@hitachi.com}
\affiliation{%
  \institution{Hitachi Power Solutions Co., Ltd.}
  \city{Hitachi}
  \state{Ibaraki}
  \country{Japan}
}


\begin{abstract}
Industrial B2B applications, such as construction site risk prediction and specialized material procurement, are characterized by extreme data sparsity yet feature rich textual information within each interaction. In such environments, traditional ID-based collaborative filtering fails due to the lack of co-occurrence signals. Conversely, fine-tuning standard Large Language Models (LLMs) presents significant practical barriers, specifically regarding high operational costs and the challenge of managing frequent data drift.

To address these challenges, we propose LLMAR (LLM-Annotated Recommendation), a tuning-free recommendation framework. Our approach is novel in that it goes beyond simple embedding utilization; by systematically integrating the reasoning capabilities of LLMs, it captures user "latent motives" without the need for a training process. Specifically, we introduce three core technical contributions: (1) Inference-Driven Annotation, which employs LLMs to transform user behavioral history from simple time-series data into structured semantic motives, enabling reasoning-based matching unattainable by ID-based methods; (2) a Reflection Loop, a self-correction mechanism where the LLM critically verifies and refines its own generated search queries to mitigate hallucinations and resolve "context competition" between past history and current instructions; and (3) a Cost-Effective Architecture, composed entirely of tuning-free components designed for asynchronous batch processing to minimize maintenance costs.

We evaluated our method using public benchmarks (MovieLens-1M, Amazon Prime Pantry) and a sparse, text-rich industrial dataset (construction site risk prediction). Experimental results demonstrate that LLMAR outperforms state-of-the-art learning-based models (SASRecF), achieving up to a 54.6\% improvement in nDCG@10 on the industrial dataset. Furthermore, the inference cost remains within a practical range of approximately \$1 per 1,000 users. Our findings suggest that in B2B domains where strict real-time latency is not critical, combining LLM reasoning and self-verification offers a superior alternative to traditional training-based approaches in terms of accuracy, explainability, and operational cost.
\end{abstract}





\maketitle

\blfootnote{Accepted at PILA '26: Workshop on Personal Intelligence in
the Agentic AI Era, co-located with ACM SIGKDD KDD 2026,
Jeju, Republic of Korea, August 10, 2026.}

\section{Introduction}

\begin{figure*}[t]
  \centering
  \includegraphics[width=\linewidth]{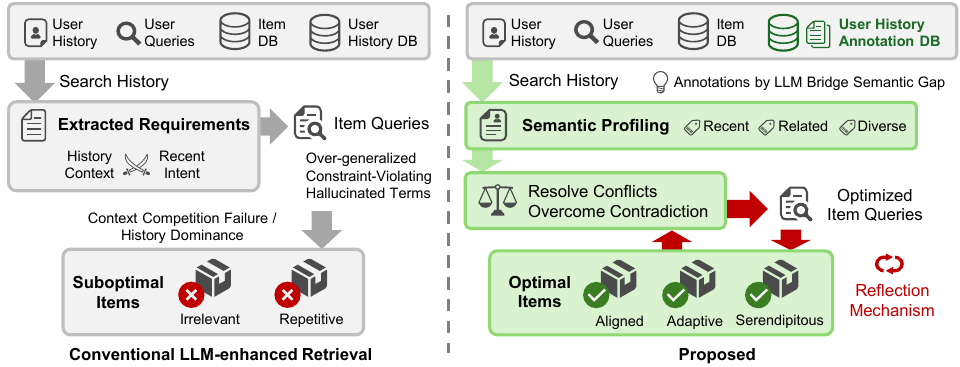}
  \caption{Comparison between conventional LLM-enhanced retrieval and the proposed framework. Conventional approaches (left) often suffer from context competition—leading to irrelevant or repetitive results due to history dominance and hallucinations. Our proposed framework (right) addresses these issues by leveraging an LLM-driven Annotation DB to bridge the semantic gap and employing a Reflection Loop to resolve conflicts, generating optimized queries to retrieve items that are aligned, adaptive, and serendipitous.}
  \label{fig:problem}
\end{figure*}

While recommender systems have achieved remarkable success in Business-to-Consumer (B2C) domains such as e-commerce and content streaming, industrial Business-to-Business (B2B) applications—including risk prediction at construction sites, specialized material procurement, and corporate matching—still face formidable challenges. In these domains, pronounced "data sparsity" due to lower total interaction volumes causes traditional ID-based Collaborative Filtering (CF) methods to perform poorly due to a lack of co-occurrence information \cite{10.1145/3038912.3052569, 10.1145/3695719.3695727}. However, a distinct characteristic of B2B domains is that individual items and interactions possess rich, specialized textual information. Therefore, the success of industrial recommender systems hinges on effectively integrating semantic text interpretation to complement sparse ID information.

Recently, the emergence of Large Language Models (LLMs), possessing extensive world knowledge and reasoning capabilities, has opened new avenues for this challenge \cite{fi17060252, 10.1145/3678004, peng-etal-2025-survey}. LLMs demonstrate superior zero-shot task performance \cite{wei2022finetunedlanguagemodelszeroshot, sun-etal-2023-chatgpt, kocmi-federmann-2023-large}, rapidly advancing their integration into recommender systems across three general paradigms \cite{10.1007/s11280-024-01291-2}. The first is the Generative Recommender \cite{10.1145/3523227.3546767, rajput2023recommender, zheng2024adapting, 10.1145/3705328.3748068}, which integrates collaborative filtering with world knowledge in the parameter space by leveraging pre-trained LLM capabilities. The second is LLM as Ranker \cite{bao2023tallrec, gao2025llm4rerank, hou2024large, 10.1145/3705328.3748055}, where candidates are initially narrowed by robust, lightweight models (e.g., Matrix Factorization \cite{10.1145/3038912.3052569}, Sequential Recommendation \cite{10.1145/3695719.3695727, 10.1145/3357384.3357895}, or GNNs \cite{10.1145/3397271.3401063}), followed by LLM reranking for flexible, prompt-based control. The third, our focus, is LLM-enhanced Retrieval \cite{li2023gpt4rec, bao2025bi, han-etal-2025-rethinking, 10.1145/3746252.3761369}. Utilizing frameworks like Retrieval-Augmented Generation (RAG) and Agents, it retrieves items via LLM-generated queries or embeddings, focusing on extracting semantic preferences through natural language profiles instead of ID-based representations \cite{10.1145/3746252.3761369, 10.1145/3711896.3737024}.

However, from a practical industrial perspective, the first two paradigms face structural barriers. Generative Recommenders require domain-specific fine-tuning, which consumes vast resources and creates operational bottlenecks due to retraining costs against daily data drift. Furthermore, tightly coupling foundation and recommendation models hinders leveraging rapidly evolving LLM capabilities. Meanwhile, LLM as Ranker struggles to process entire item pools due to computational constraints; because it relies on upstream retrieval performance, it cannot fundamentally improve initial retrieval accuracy.

In contrast, LLM-enhanced Retrieval allows for a tuning-free configuration, offering a superior cost-flexibility balance. Nevertheless, simply vectorizing items is insufficient for complex B2B decision-support scenarios. Two essential challenges remain. First is the semantic gap between behavioral history and recommended items. B2B user actions stem from deep "latent motives" unreadable from superficial logs. While existing work \cite{han-etal-2025-rethinking} infers preferences from history, an improved process is needed to interpret temporal preference changes and convert them into explicit "Search Intent" for future items. Second is context competition. Implicit constraints from past behavior often conflict with explicit constraints like current user instructions or diversity needs. In high-reliability tasks, inappropriate recommendations from past trends or LLM-specific hallucinations are unacceptable. Addressing this requires models to critically verify and correct their outputs iteratively \cite{asai2023selfraglearningretrievegenerate, yao2023reactsynergizingreasoningacting}.

Our approach explicitly addresses these challenges. First, we build a semantic retrieval infrastructure by using an LLM to generate and embed item text. Next, to bridge the semantic gap, we treat user history as time-series "bundles" and use an LLM to accumulate "Annotations" that extract preference evolution. During recommendation, these Annotations generate search queries, projecting latent preferences into the search space. Furthermore, we introduce a Reflection Loop to resolve context competition. This allows the LLM to verify and correct whether generated queries and results contradict current intent, balancing diversity and accuracy. As shown in Fig.~\ref{fig:problem}, unlike conventional methods where fragmented preferences cause context contradictions, our proposed framework utilizes Annotation and Reflection mechanisms. Because all components use tuning-free LLMs, the framework is highly advantageous for model updates and seamlessly handles arbitrary natural language inputs at runtime.

The main contributions of this study are as follows:

\begin{itemize}
 \item We propose a novel tuning-free recommendation framework that resolves the semantic gap and context competition through LLM reasoning capabilities (Annotation and Reflection).
 \item The proposed method demonstrates competitive accuracy against learning-based methods on general open datasets, achieving up to a 54.6\% improvement in nDCG@10 over conventional models on industrial datasets.
 \item We demonstrate low-cost operational feasibility (on the order of \$1 per 1,000 users) and verify explainability through qualitative analysis, proving it a compelling alternative to traditional learning-based approaches in sparse, text-rich B2B domains.
\end{itemize}
\section{Related Work}

\subsection{LLM-Integrated Recommender Systems}
The integration of LLMs into recommender systems has primarily advanced along three trajectories to bridge the semantic gap inherent in conventional ID-based methods: Generative Recommenders, LLM as Ranker, and LLM-enhanced Retrieval.

Generative Recommenders \cite{10.1145/3523227.3546767, rajput2023recommender, zheng2024adapting} treat item IDs as language tokens and formulate the recommendation task as a sequence generation problem. Represented by models such as P5 \cite{10.1145/3523227.3546767}, these approaches demonstrate high performance through multi-task learning. However, large-scale fine-tuning is indispensable to align the ID space with the linguistic space. This entails high operational costs, particularly in addressing the cold-start problem and in real-world environments requiring frequent item updates.
On the other hand, the LLM as Ranker paradigm \cite{bao2023tallrec, hou2024large, gao2025llm4rerank} leverages the reasoning capabilities of LLMs specifically for re-ranking candidate items. Nevertheless, due to inference costs and context window constraints, scoring the entire item pool is infeasible; thus, performance remains heavily dependent on the precision of the preceding lightweight retriever.

In contrast, this study focuses on LLM-enhanced Retrieval \cite{li2023gpt4rec, bao2025bi, han-etal-2025-rethinking}, which utilizes LLMs as encoders or query generators to enable full-item retrieval while curbing computational costs.
Notably, Han et al.~\cite{han-etal-2025-rethinking} attempt to overcome the limitations of ID-based methods by describing user behavioral history in natural language and performing retrieval using its embedding representation.
However, existing RAG-based methods are limited to either concatenating behavioral history into a single long sequence or summarizing it into a static profile \cite{wang2023zeroshotnextitemrecommendationusing, 10.1145/3616855.3635845}.
These approaches do not adequately address "Temporal Granularity"—specifically, how user preferences evolve over time—and a deeper investigation into the transition and invariance of preferences remains inadequately explored.
This study addresses this information loss by structuring history into chronological "Bundles" and assigning LLM-generated annotations.

\subsection{Reasoning and Self-Correction in LLMs}
Approaches to enhance the reasoning capabilities of LLMs have been established through Chain-of-Thought (CoT) \cite{10.5555/3600270.3602070} and Reflection (or Self-Correction) mechanisms, where the model critically verifies and corrects its own output \cite{asai2023selfraglearningretrievegenerate, yao2023reactsynergizingreasoningacting, 10.5555/3666122.3666499}.
These techniques effectively mitigate hallucinations and improve logical consistency in tasks requiring complex reasoning by interposing intermediate thought processes.

In the domain of recommender systems, attempts have been made to introduce reasoning steps within agent-based models \cite{10.1145/3705328.3748055, 10.1609/aaai.v39i12.33434}.
However, most prior works have focused on response generation in conversational interfaces or ensuring consistency in the justification of final rankings.
While Self-RAG \cite{asai2023selfraglearningretrievegenerate} and ReAct \cite{yao2023reactsynergizingreasoningacting} demonstrate iterative self-correction in retrieval-augmented generation pipelines, these methods operate on the \emph{post-retrieval} or \emph{response-generation} stage.
While self-correction has proven effective in general LLM tasks, its application to the \emph{pre-retrieval query generation} stage in recommender systems remains largely unexplored. Specifically, existing literature lacks mechanisms to resolve the ``context competition'' that arises when a user's latent long-term intent conflicts with their recent behavioral history.
The Reflection Loop proposed in this study is novel in that it autonomously verifies whether the search query aligns with the user's current intent and optimizes the projection into the search space.
\section{Methodology}

\begin{figure*}[t]
    \centering
    \includegraphics[width=\linewidth]{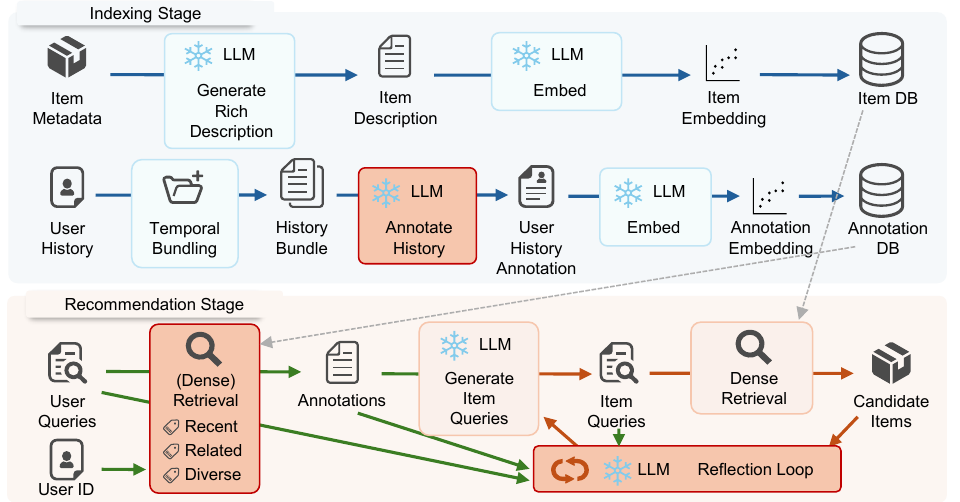}
    \caption{\textbf{The schematic overview of the LLMAR framework.} The architecture consists of two main phases: (1) the \textit{Indexing Stage}, where item metadata is enriched and user history is transformed into latent motive annotations via inference-driven reasoning; and (2) the \textit{Recommendation Stage}, which utilizes a dual-retrieval strategy to synthesize search queries based on retrieved annotations. A self-correcting \textit{Reflection Loop} dynamically verifies and refines candidate items to align with the user's potential needs.}
    \label{fig:methodology}
\end{figure*}

We introduce the LLMAR (LLM-Annotated Recommendation) framework to bridge the semantic gap between user history and queries while resolving context competition. As illustrated in Fig.~\ref{fig:methodology}, LLMAR utilizes a three-stage process: (1) Inference-Driven Annotation, (2) a Dual-Retrieval Strategy, and (3) a Reflection Loop.

\subsection{Problem Formulation}

Let $\mathcal{U}$ and $\mathcal{I}$ denote user and item sets. Item $i \in \mathcal{I}$ has metadata $M_i$ (descriptions, attributes). User $u \in \mathcal{U}$ has metadata $W_u$ (demographics) and interaction history $H_u = \{(i_1, c_1), \dots, (i_t, c_t)\}$, with $c_k$ denoting interaction context (e.g., timestamp). Our objective is to generate a recommended item list $\hat{\mathcal{I}}$ satisfying user needs based on $H_u$ and an optional input query $q$.

\subsection{Knowledge-Augmented Item Representation}

To mitigate cold-start and semantic sparsity issues in ID-based methods, we perform offline LLM-based item augmentation. Item metadata $M_i$ is fed into an LLM to generate a rich natural language description $D_i$ incorporating contextual background:
\begin{equation}
D_i \sim p_{\theta_{\text{LLM}}}(D \mid M_i, P_{\text{item}})
\end{equation}
where $P_{\text{item}}$ prompts for searchable appeal. We concatenate $D_i$ with $M_i$ to form augmented metadata $M^+_i = [M_i; D_i]$, yielding vector embeddings $\mathbf{e}_i = E(D_i)$ via embedding model $E(\cdot)$. This enables semantic retrieval combining factual attributes with LLM-inferred contextual nuances.

\subsection{Step 1: Inference-Driven Annotation}

We transform surface-level interaction histories into user ``latent motives,'' extracting sequential intents to articulate preferences and enhance noise robustness.

\subsubsection{Bundle Construction and Reasoning}

Because user preferences evolve over time, we partition history $H_u$ into $m$ context-aware ``bundles'' $\mathcal{S} = \{S_1, \dots, S_m\}$ via temporal sliding windows. For each $S_j$, an LLM infers an underlying common selection motive $r_j$:
\begin{equation}
r_j \sim p_{\theta_{\text{LLM}}}(r \mid P_{\text{annotate}}, \{M^+_i \mid i \in S_j\}, W_u)
\end{equation}
where $P_{\text{annotate}}$ prompts for preference articulation. This abstracts factual enumerations (e.g., ``watched Movie A and B'') into searchable intents (e.g., ``preference for Sci-Fi with complex plots''), redefining user history as a motive-annotated profile $H'_u = \{(S_j, r_j)\}_{j=1}^m$.

\subsubsection{Motive Vectorization}

Extracted motives $r_j$ are embedded into a vector set $V_u$:
\begin{equation}
V_u = \{ \mathbf{v}_j \mid \mathbf{v}_j = E(r_j), j=1 \dots m \}
\end{equation}
This enables preference computation within a continuous semantic space rather than a discrete ID space.

\subsection{Step 2: Dual-Retrieval Strategy}

Standard similarity retrieval misses latent need diversity. We thus synthesize queries by retrieving motives via ``Exploitation'' (deepening known interests) and ``Exploration'' (broadening unknown interests).

\subsubsection{Motive Retrieval}

Given input $q$, we compile a candidate motive set $R_{\text{total}}$ via two strategies:

\paragraph{Exploitation (History Matching):}
We retrieve motives from $V_u$ matching explicit instructions or recent history, ensuring preference consistency:
\begin{equation}
R_{\text{exploit}} = 
\begin{cases} 
\text{TopK}_{\mathbf{v} \in V_u}(\text{sim}(\mathbf{v}, E(q))) & \text{if } q \neq \emptyset \\ 
\text{Latest}_k(V_u) & \text{if } q = \emptyset 
\end{cases}
\end{equation}

\paragraph{Exploration (Diversity \& Serendipity):}
To mitigate filter bubbles, we extract diverse motives $R_{\text{explore}}$ via:

\begin{itemize}
    \item \textit{Intra-user Diversity (MMR):} We maximize relevance while reducing redundancy within $V_u$ using Maximal Marginal Relevance (MMR) \cite{10.1145/290941.291025}:
    \begin{equation}
    \mathbf{v}^* = \underset{\mathbf{v} \in V_u \setminus R_{\text{exploit}}}{\text{argmax}} \left[ \lambda \text{sim}(\mathbf{v}, E(q)) - (1-\lambda) \max_{\mathbf{v}' \in R_{\text{exploit}}} \text{sim}(\mathbf{v}, \mathbf{v}') \right]
    \end{equation}
    The penalty term minimizes overlap with $R_{\text{exploit}}$. Iterating $k$ times yields a diverse set $R_{\text{div}}$.
    
    \item \textit{Social Expansion (Collaborative Retrieval):} We identify similar motives from other users' pools $V_{\text{others}}$:
    \begin{equation}
    R_{\text{social}} = \text{TopK}_{\mathbf{v} \in V_{\text{others}}}(\text{sim}(\mathbf{v}, E(q)))
    \end{equation}
    This semantic collaborative filtering infers unarticulated latent needs from peer behaviors.
\end{itemize}

The final motive set is $R_{\text{total}} = R_{\text{exploit}} \cup R_{\text{div}} \cup R_{\text{social}}$.

\subsubsection{Query Synthesis}

Since $R_{\text{total}}$ may contain conflicting motives, an LLM integrates these to synthesize executable search queries:
\begin{equation}
Q_{\text{search}} \sim p_{\theta_{\text{LLM}}}(Q \mid q, R_{\text{total}}, P_{\text{query}})
\end{equation}
To minimize latency, $P_{\text{query}}$ prompts directly for search-engine-optimized queries without intermediate reasoning steps.

\subsection{Step 3: Iterative Search \& Verification}

\subsubsection{Search and Ranking}

LLM-generated queries may yield suboptimal results due to context competition. We search using $Q_{\text{search}}$ to yield candidates $\mathcal{C}$, then integrate multiple query results via Reciprocal Rank Fusion (RRF) \cite{10.1145/1571941.1572114} for robust ranking:
\begin{equation}
\mathcal{C}_{\text{ranked}} = \text{RRF}(\{\text{Retrieve}(\mathcal{I}, q') \mid q' \in Q_{\text{search}}\})
\end{equation}

\subsubsection{Reflection Loop}

The LLM evaluates the validity of $\mathcal{C}_{\text{ranked}}$ and refines queries:
\begin{equation}
(\text{Score}, Q'_{\text{search}}) \sim p_{\theta_{\text{LLM}}}(\cdot \mid \mathcal{C}_{\text{ranked}}, q, R_{\text{total}}, Q_{\text{search}}, P_{\text{reflect}})
\end{equation}
If the score meets threshold $\tau$ or max iterations $T_{\max}$ are reached, the loop terminates; otherwise, re-retrieval occurs with updated $Q'_{\text{search}}$. This Reflection Loop dynamically mitigates hallucinations and intent-result discrepancies (Algorithm \ref{alg:llmar}).

\begin{algorithm}[t]
\caption{LLMAR Inference Process}
\label{alg:llmar}
\begin{algorithmic}[1]
\Require User history bundles $V_u$, Current query $q$ (optional), Global motive pool $V_{\text{others}}$
\Ensure Recommended item list $\mathcal{C}_{\text{ranked}}$

\State \Comment{Step 1: Motive Retrieval}
\If{$q \neq \emptyset$}
    \State $R_{\text{exploit}} \leftarrow \text{TopK}(\text{sim}(V_u, E(q)))$
\Else
    \State $R_{\text{exploit}} \leftarrow \text{Latest}(V_u)$
\EndIf
\State $R_{\text{div}} \leftarrow \text{MMR}(V_u, E(q), R_{\text{exploit}})$
\State $R_{\text{social}} \leftarrow \text{TopK}(\text{sim}(V_{\text{others}}, E(q)))$
\State $R_{\text{total}} \leftarrow R_{\text{exploit}} \cup R_{\text{div}} \cup R_{\text{social}}$

\State \Comment{Step 2: Query Synthesis}
\State $Q_{\text{search}} \leftarrow \text{LLM}_{\text{gen}}(q, R_{\text{total}})$

\State \Comment{Step 3: Iterative Search \& Verification}
\For{$t = 1$ to $T_{\max}$}
    \State $\mathcal{C} \leftarrow \text{DenseRetrieval}(\mathcal{I}, Q_{\text{search}})$
    \State $\mathcal{C}_{\text{ranked}} \leftarrow \text{RRF}(\mathcal{C})$
    \State $(\text{Score}, \text{Feedback}) \leftarrow \text{LLM}_{\text{verify}}(\mathcal{C}_{\text{ranked}}, q, R_{\text{total}})$
    
    \If{$\text{Score} \geq \tau$}
        \State \textbf{break}
    \EndIf
    \State $Q_{\text{search}} \leftarrow \text{UpdateQuery}(Q_{\text{search}}, \text{Feedback})$
\EndFor

\State \Return $\mathcal{C}_{\text{ranked}}$
\end{algorithmic}
\end{algorithm}
\section{Experiments}

In this section, we conduct experiments to answer the following four research questions (RQs) to comprehensively evaluate the effectiveness of the proposed method, LLMAR.

\begin{itemize}
    \item RQ1 (Overall Performance): Does the proposed method demonstrate practical utility in terms of recommendation accuracy and diversity compared to state-of-the-art (SOTA) ID-based models?
    \item RQ2 (Ablation Study): How does each proposed module (Annotation, Exploration, and Reflection) contribute to the final performance?
    \item RQ3 (Qualitative Analysis): Does LLMAR correctly interpret users' ``latent motives'' and appropriately resolve competing contexts? We verify whether the process possesses explainability and transparency through specific case studies.
    \item RQ4 (Efficiency \& Limitations): Are the inference cost and latency within a practical range?
\end{itemize}

\subsection{Experimental Setup}

\subsubsection{Datasets}
We selected three datasets with varying data densities and domains (Table \ref{tab:datasets}).

\begin{enumerate}
    \item MovieLens-1M (ML-1M) \cite{10.1145/2827872}: A benchmark dominated by high-density, ID-based signals. It contains abundant categorical variables as metadata. This dataset is used to verify the baseline performance of the proposed method in an environment where traditional collaborative filtering is typically advantageous.
    \item Amazon Prime Pantry \cite{10.1145/2872427.2883037, 10.1145/2766462.2767755}: A dataset characterized by a fast consumption cycle, a vast number of items, and high sparsity. Abundant categorical metadata is available. Semantic understanding of item metadata is more critical than ID co-occurrence information.
    \item Industrial Safety Risk (ISR): A proprietary industrial dataset for risk prediction in construction and maintenance sites. While less sparse than Amazon Pantry (Density: 1.49\%), ISR remains significantly sparser than typical B2C platforms and is too sparse for effective collaborative filtering. Each risk instance is annotated with semantic text descriptions such as risk types and expected accident categories. This dataset plays a crucial role in verifying the effectiveness of semantic interpretation by LLMs. The problem is defined as a sequential recommendation task that suggests new relevant risks to on-site users based on historical risk analysis logs.
\end{enumerate}

\begin{table}[t]
    \centering
    \caption{Statistics of the datasets used in the experiments. We applied 5-core filtering and retained only interactions with ratings of 3 or higher.}
    \label{tab:datasets}
    \small % reduce font size to conserve space
    \begin{tabular}{lcccc}
        \toprule
        Dataset & \# Users & \# Items & \# Interactions & Density \\
        \midrule
        MovieLens-1M & 6,039 & 3,308 & 835,789 & 4.18\% \\
        Amazon Pantry & 12,856 & 4,724 & 125,387 & 0.206\% \\
        ISR (Risk) & 1,340 & 2,340 & 46,800 & 1.49\% \\
        \bottomrule
    \end{tabular}
\end{table}

\subsubsection{Baselines}
To ensure a fair comparison, we adopted strong baselines representing each category.
\begin{itemize}
    \item LightGCN \cite{10.1145/3397271.3401063}: A SOTA model for collaborative filtering using Graph Convolutional Networks.
    \item SASRec \cite{10.1145/3695719.3695727}: A SOTA model for sequential recommendation using Self-Attention.
    \item SASRecF: An extended model of SASRec that integrates item attributes (metadata) as features. Since it can utilize meta-information, it serves as the most direct baseline for comparison in this study.
\end{itemize}

\subsubsection{LLM Setup}
We used Qwen3-235B-A22B-Instruct-2507 \cite{qwen3technicalreport} for text generation tasks such as Annotation and query generation. For natural language embeddings, we utilized Cohere's Embed Multimodal v4.0 \cite{cohere2025embedv4}. All models were accessed via the Amazon Bedrock API \cite{aws_bedrock}.

\subsubsection{Evaluation Protocol \& Reproducibility}
To ensure evaluation fairness, we adopted a timestamp-based chronological split (8:1:1) for all datasets. For ranking evaluation, we performed a full ranking on all candidate items to eliminate bias caused by sampling (e.g., extracting only 100 negative samples). We used Recall@K, nDCG@K, and MRR@K as accuracy metrics, and Coverage@K and Popularity@K (average popularity) as diversity metrics. To ensure reproducibility, the code and parameter settings used for the baselines and the proposed method will be made available in a GitHub repository\footnote{https://github.com/hishikawa-hitachi/kdd-pila-2026-submission-code}. We used the well-verified Python library RecBole\footnote{https://github.com/RUCAIBox/RecBole} \footnote{https://github.com/RUCAIBox/RecSysDatasets} \cite{recbole[1.2.1]} for evaluating the baselines.

\subsection{Performance Comparison (RQ1)}

The performance comparison for each dataset is shown in Table \ref{tab:performance}.

\begin{table*}[t]
    \centering
    \caption{Performance Comparison on Three Datasets. Bold indicates the best result, underlined value indicates the second best.}
    \label{tab:performance}
    \resizebox{\textwidth}{!}{%
    \begin{tabular}{llcccccccccc}
        \toprule
        Dataset & Model & Recall@10 & Recall@20 & nDCG@10 & nDCG@20 & MRR@10 & MRR@20 & Cov@10 & Cov@20 & Pop@10 & Pop@20 \\
        \midrule
        \multirow{4}{*}{\textbf{MovieLens-1M}} & LightGCN & 0.0674 & 0.1130 & 0.0758 & 0.0873 & \textbf{0.1412} & \textbf{0.1506} & 0.3510 & 0.4680 & 1304.21 & 1174.89 \\
         & SASRec & \underline{0.2089} & \underline{0.3064} & \underline{0.1072} & \underline{0.1318} & 0.0764 & 0.0831 & \underline{0.9284} & \underline{0.9664} & 549.36 & 533.60 \\
         & SASRecF & \textbf{0.2125} & \textbf{0.3093} & \textbf{0.1131} & \textbf{0.1375} & \underline{0.0831} & \underline{0.0897} & 0.9144 & 0.9583 & \underline{548.65} & \underline{531.49} \\
         & LLMAR (Ours) & 0.1942 & 0.2883 & 0.1015 & 0.1257 & 0.0721 & 0.0792 & \textbf{0.9414} & \textbf{0.9753} & \textbf{485.30} & \textbf{462.19} \\
        \midrule
        \multirow{4}{*}{\textbf{Amazon Pantry}} & LightGCN & 0.0357 & 0.0526 & 0.0229 & 0.0272 & 0.0193 & 0.0206 & 0.8385 & \underline{0.9369} & 120.22 & 107.48 \\
         & SASRec & \textbf{0.0900} & \textbf{0.1180} & \underline{0.0578} & \underline{0.0648} & \underline{0.0478} & \underline{0.0497} & \underline{0.8707} & 0.9124 & 91.64 & 85.96 \\
         & SASRecF & \underline{0.0859} & \underline{0.1057} & \textbf{0.0619} & \textbf{0.0669} & \textbf{0.0546} & \textbf{0.0559} & 0.7756 & 0.8683 & \underline{80.98} & \underline{75.56} \\
         & LLMAR (Ours) & 0.0812 & 0.1012 & 0.0545 & 0.0610 & 0.0454 & 0.0482 & \textbf{0.8952} & \textbf{0.9421} & \textbf{72.42} & \textbf{68.12} \\
        \midrule
        \multirow{4}{*}{\textbf{ISR (Risk)}} & LightGCN & 0.0285 & 0.0412 & 0.0165 & 0.0205 & 0.0112 & 0.0135 & 0.3250 & 0.4120 & 95.12 & 88.40 \\
         & SASRec & 0.0682 & 0.0955 & 0.0385 & 0.0452 & 0.0315 & 0.0340 & 0.6540 & 0.7210 & 68.45 & 64.20 \\
         & SASRecF & \underline{0.0785} & \underline{0.1082} & \underline{0.0445} & \underline{0.0528} & \underline{0.0368} & \underline{0.0395} & \underline{0.7120} & \underline{0.7850} & \underline{62.30} & \underline{58.15} \\
         & LLMAR (Ours) & \textbf{0.1215} & \textbf{0.1651} & \textbf{0.0688} & \textbf{0.0811} & \textbf{0.0552} & \textbf{0.0598} & \textbf{0.9258} & \textbf{0.9687} & \textbf{48.11} & \textbf{45.34} \\
        \bottomrule
    \end{tabular}
    }
\end{table*}

\subsubsection{Robustness in Text-Rich Sparse Scenarios}
The results on ISR and Amazon Pantry demonstrate the competitiveness and robustness of the proposed method. Notably, on ISR, LLMAR achieved a significant performance improvement over SASRecF, with +54.8\% in Recall@10 and +54.6\% in nDCG@10. Traditional ID-based methods struggle to acquire sufficient item representations due to data sparsity, resulting in poor generalization to unknown contexts. In contrast, LLMAR leverages the world knowledge inherent in LLMs to perform ``semantic matching'' beyond superficial text matches, successfully capturing the user's (site's) intent even from few interactions.

\subsubsection{Competitiveness in Dense Scenarios}
On high-density datasets like MovieLens-1M, LLMAR showed performance comparable to the learning-based SOTA model (SASRecF). Since the task involves future prediction using chronological splitting rather than random splitting, sequential models perform well. Although SASRecF numerically outperforms LLMAR, this is attributed to the extreme power of collaborative filtering when dense ID co-occurrence information is available. Notably, LLMAR achieved this accuracy without any parameter updates. In environments where the cold-start problem exists or frequent model updates are difficult, LLMAR offers a new optimal solution to the cost-accuracy trade-off.

\subsubsection{Diversity and Bias Mitigation}
Across all datasets, LLMAR recorded the highest Coverage and the lowest Popularity Bias (Pop@10). This suggests it avoids the bias of traditional methods that tend to learn popular items, successfully discovering useful items in the long tail. In particular, ``Social Expansion'' (borrowing motives from other users) in the Exploration module likely contributed to serendipitous recommendations.

\subsection{Ablation Study (RQ2)}

We conducted an ablation study on the ML-1M dataset to confirm the contribution of each component of the proposed method (Table \ref{tab:ablation}).

\begin{enumerate}
    \item w/o Annotation: Disables the first stage of interaction history inference and uses only summaries.
    \item w/o Exploration: Disables the second stage of exploratory search and uses only highly similar motives (exploitation only).
    \item w/o Reflection: Disables the third stage (self-verification loop) and outputs the result in a single pass.
\end{enumerate}

\begin{table*}[t]
    \centering
    \caption{Ablation Study Results on MovieLens-1M}
    \small
    \label{tab:ablation}
    \begin{tabular}{lcc}
        \toprule
        Variant & nDCG@10 (Performance Drop) & Pop@10 (Performance Drop) \\
        \midrule
        \textbf{Full LLMAR} & \textbf{0.1015} & \textbf{485.30} \\
        w/o Annotation & 0.0776 (-23.5\%) & 523.15 (+7.80\%) \\
        w/o Exploration & 0.0915 (-9.82\%) & 785.22 (+61.8\%) \\
        w/o Reflection & 0.0626 (-38.3\%) & 685.73 (+41.3\%) \\
        \bottomrule
    \end{tabular}
\end{table*}

\begin{itemize}
    \item Effect of Reflection: Removing the verification loop caused the largest drop in accuracy (-38.3\%). This indicates that the Reflection mechanism is indispensable as a ``quality guardian'' to filter out hallucinations and inappropriate search results.
    \item Effect of Annotation: Removing annotations also resulted in performance degradation. This suggests that ``latent motives'' extracted by the LLM are more effective than simple summaries in bridging the semantic gap between interaction history and recommended items.
    \item Effect of Exploration: Removing the exploration module had a limited impact on accuracy but significantly worsened popularity bias (+61.8\%). This indicates that this module is the primary factor in preventing filter bubbles and ensuring diversity.
\end{itemize}

\subsection{Qualitative Analysis: Solving Context Competition (RQ3)}

\begin{figure*}[t]
    \centering
    \includegraphics[width=\linewidth]{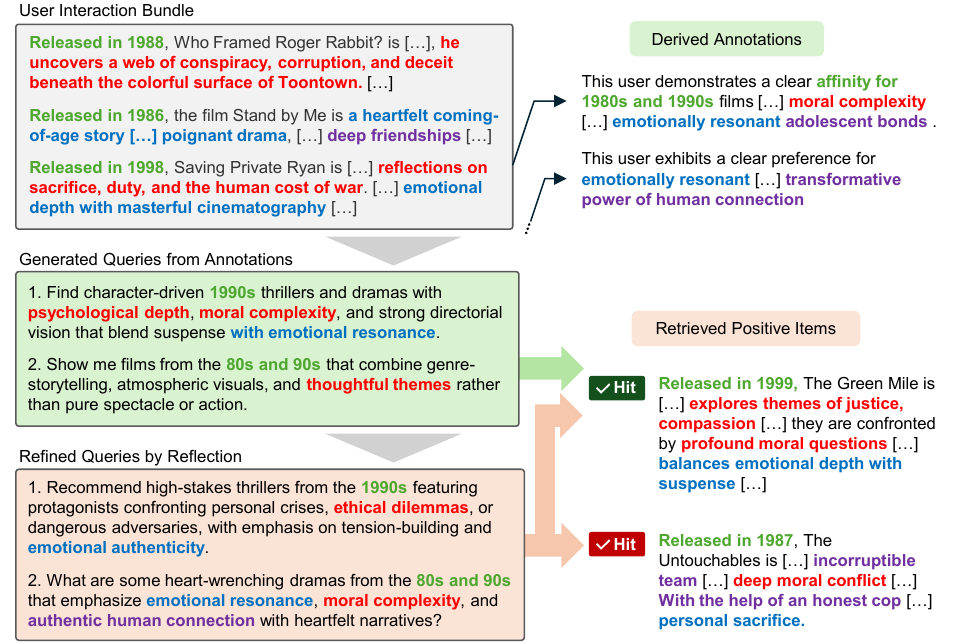}
    \caption{Qualitative analysis of the inference process. The framework utilizes Derived Annotations to capture latent user motives (e.g., "moral complexity") from interaction bundles, bridging the semantic gap. Subsequently, the Reflection mechanism refines the initial queries into more precise forms to ensure alignment with the target items. Note: For clarity, only representative excerpts of interactions, queries, and annotations are displayed.}
    \label{fig:qualitative_analysis}
\end{figure*}

In this section, we analyze the actual inference process (Figure \ref{fig:qualitative_analysis}) to verify how LLMAR bridges the ``semantic gap'' between interaction history and recommendation candidates and resolves ``context competition.''

\subsubsection{Bridging the Semantic Gap via Inference-Driven Annotation}
In traditional ID-based or simple keyword matching methods, seemingly inconsistent interaction histories are often treated as noise. In the case shown in the figure, the target user's history includes \textit{Who Framed Roger Rabbit} (Comedy/Crime), \textit{Stand by Me} (Drama), and \textit{Saving Private Ryan} (War), which are disparate in terms of genre. However, the Annotation module of our method successfully extracts ``latent motives'' that transcend these superficial categories. Specifically, it identified common thematic structures such as ``moral complexity'' and the ``transformative power of human connection'' from individual descriptions. This demonstrates that the proposed method interprets items not merely as sets of attributes but as contexts possessing ``Narrative Depth,'' defining the user's fundamental preferences within a semantic space. Indeed, in these results, the item search query generated from the Annotation successfully retrieved the ground truth, \textit{The Green Mile}.

\subsubsection{Resolving Context Competition via Reflection Loop}
The most crucial finding in this case study is the resolution of ``context competition'' via the Reflection Loop. In the initial query generation stage (Generated Queries), the model was biased toward the strong context of the most recent viewing history, \textit{Saving Private Ryan} (War, Suspense), forming search intents biased toward ``Dark Thriller'' and ``Psychological Depth.'' This is a typical case of context competition (History Dominance), where recent strong signals overwrite long-term preferences (the ``Heartfelt'' elements seen in \textit{Stand by Me}).

In response, the Reflection Loop verified the consistency between the generated queries and the accumulated Annotations, detecting this bias. Consequently, the Refined Queries mediated the conflicting constraints by explicitly adding ``heart-wrenching'' and ``authentic human connection'' to the search scope while maintaining ``moral complexity.'' This self-correction process functions to prevent ``filter bubbles'' and ``excessive bias''—which are difficult to avoid in single-pass inference—and dynamically recovers the diversity of the search space.

\begin{table*}[t]
    \centering
    \caption{Inference Latency and Cost on MovieLens-1M.}
    \label{tab:backbone}
    \small
    \begin{tabular}{lccc}
        \toprule
        Backbone & nDCG@10 & Latency (Initial / +Reflect) & Cost per 1k Users (Initial / +Reflect)\\
        \midrule
        Qwen3 235B A22B & 0.1015 & 2.0s / 3.4s & ~\$0.9 / ~\$1.7 \\
        Qwen3 Next 80B A3B & 0.0875 & 1.4s / 2.6s & ~\$0.7 / ~\$1.6 \\
        \bottomrule
    \end{tabular}
\end{table*}

\subsubsection{Effectiveness of Motive-Oriented Retrieval}
The final search results demonstrate semantic alignment beyond keyword matching. The recommended \textit{The Green Mile} (1999), while technically Fantasy/Crime, fully aligns with the motive of ``exploring themes of justice'' extracted from the user's history. Similarly, \textit{The Untouchables} (1987) functions as a bridge between the ``conspiracy and corruption'' elements of the comedy \textit{Roger Rabbit} and the serious moral conflicts of the war movie.

From this analysis, we conclude that LLMAR achieves not merely superficial matching of ``What'' (what was watched), but an interpretation of the underlying ``Why'' (why it was watched) and a dynamic optimization of the ``How'' (how to search). This interpretability and reasoning capability are the factors supporting the quantitative superiority in sparse environments confirmed in the previous section (Table~\ref{tab:performance}).

\subsection{Efficiency \& Limitations (RQ4)}
Finally, we discuss computational costs and the limitations of this method.

\subsubsection{Computational Cost}
As shown in Table \ref{tab:backbone}, inference time increases with the size of the LLM backbone. A latency of several seconds per user may not be suitable for real-time B2C web services requiring millisecond-level responses. However, for scenarios like the industrial risk prediction (ISR) targeted by this method or periodic purchasing in Amazon Pantry, immediacy is not necessarily required. For instance, by adopting implementation forms such as overnight batch processing for next-day risk prediction or asynchronous recommendation list updates, this computational cost is considered well within the acceptable range for practical operation. Furthermore, this method is optimal for B2B services with fewer users, such as industrial risk prediction, where accuracy and log interpretability/transparency are demanded. Additionally, when using LLMs via APIs like Amazon Bedrock, asynchronous calls are standard, making item recommendation completely parallelizable from the recommendation system server's perspective. Therefore, scalability relative to the number of users is extremely high, and the impact of user growth on application operation is limited. The recommendation cost is on the order of \$1 per 1,000 users, which is a realistic value. This is because the text generated by the LLM is limited to item queries. The method also offers customizability; for example, depending on the allowable cost and latency, employing Chain-of-Thought to output the reasoning process could further improve accuracy.

\subsubsection{Failure Cases}
Analysis revealed that a typical failure case for LLMAR is the recommendation of ``items with extremely poor metadata.'' Since the LLM relies on text information for inference, it struggles to perform semantic matching for items lacking meaningful categorical variables other than the product name, showing a tendency to underperform compared to ID-based methods. Furthermore, when multiple items in the pool share nearly identical metadata—such as product name, category, rating, and timestamp—semantic differentiation becomes difficult. In such cases, similar items dominate the top rankings, leading to a phenomenon where semantic diversity (which does not appear in general metrics like Average Popularity) decreases.
\section{Conclusion and Future Work}

In this study, we proposed LLMAR, a tuning-free framework that systematically integrates LLM reasoning capabilities to address "semantic gaps between behavioral history and recommendation candidates" and "context competition" in B2B applications characterized by data sparsity and rich text.
Through extensive experiments across three distinct domains and qualitative analysis, our approach provides the following critical implications for real-world deployment:

First, we demonstrated the \textbf{superiority of a reasoning-based approach in sparse data environments}.
In industrial datasets (ISR) where "co-occurrence" signals required by traditional ID-based collaborative filtering are scarce, LLMAR drastically outperformed state-of-the-art learning-based models (+54.6\% improvement in nDCG@10).
This proves that in metadata-rich domains, interpreting user "Motives" via LLMs serves as a robust alternative for overcoming cold-start and data sparsity problems.

Second, the framework achieves \textbf{a balance between explainability and controllability}.
The Reflection Loop functions as a "quality guardian," suppressing hallucinations while maintaining consistency among potentially conflicting historical interactions and shifting preferences.
Furthermore, natural language annotations generated by LLMAR bring transparency to recommendation reasoning—conventionally a black box—providing significant practical value in B2B applications where reliability is paramount.

Third, we observed \textbf{substantial improvements in operational cost and maintainability}.
Because LLMAR consists entirely of tuning-free components, it eliminates major MLOps barriers like adapting to frequent data drift and model retraining.
With inference costs at approximately \$1 per 1,000 users, this is an immediately deployable solution for asynchronous batch-processing B2B applications where strict real-time constraints are not primary concerns.

Identifying technical challenges for real-world deployment has also highlighted promising directions for future research.

\textbf{Dependency on Metadata Quality:}
Our method's performance relies on textual information quality, potentially underperforming compared to ID-based methods when item descriptions are poor.
Expanding the framework to Multimodal LLMs (MLLMs) integrating non-textual modalities—such as image data and interaction logs—is a promising approach to compensate for incomplete metadata and broaden applicable domains.

\textbf{Inference Latency \& Cost:}
While multi-step LLM reasoning enhances interpretability, achieving millisecond-level latency required for B2C applications remains challenging.
Future work involves distinguishing between domain-knowledge-specific and reasoning-specific tasks during API calls. By utilizing specialized models to limit parameter counts, we expect to compress latency for real-time operations while maintaining inference quality.

\textbf{Temporal Granularity:}
Our implementation uses fixed time windows for segmenting history, yet user interests do not shift at uniform intervals.
Introducing dynamic clustering based on behavioral semantic similarity is the next step. This allows for a more precise separation of contextually distinct interactions (e.g., personal hobbies vs. gift purchases), improving the granularity of generated Annotations.

% \begin{acks}
% \end{acks}

\bibliographystyle{ACM-Reference-Format}
\bibliography{sample-base}

% \appendix

% hello

\end{document}